\makeatletter
\newcommand{\dontusepackage}[2][]{%
  \@namedef{ver@#2.sty}{9999/12/31}%
  \@namedef{opt@#2.sty}{#1}}
\makeatother
\dontusepackage{subfigure}

\documentclass[]{article}

\usepackage{lmodern}
\usepackage{amssymb,amsmath}
\usepackage{ifxetex,ifluatex}
\usepackage[usenames,dvipsnames]{color}
\usepackage{fixltx2e} 
\ifnum 0\ifxetex 1\fi\ifluatex 1\fi=0 
  \usepackage[T1]{fontenc}
  \usepackage[utf8]{inputenc}
\else 
  \ifxetex
    \usepackage{mathspec}
    \usepackage{xltxtra,xunicode}
  \else
    \usepackage{fontspec}
  \fi
  \defaultfontfeatures{Mapping=tex-text,Scale=MatchLowercase}
  
\fi
\IfFileExists{upquote.sty}{\usepackage{upquote}}{}
\IfFileExists{microtype.sty}{%
\usepackage{microtype}
\UseMicrotypeSet[protrusion]{basicmath} 
}{}
\usepackage[margin=1.0in,bottom=1.5in]{geometry}
\usepackage[square,numbers,sort&compress]{natbib}
\usepackage{listings}
\usepackage{graphicx}
\makeatletter
\def\maxwidth{\ifdim\Gin@nat@width>\linewidth\linewidth\else\Gin@nat@width\fi}
\def\maxheight{\ifdim\Gin@nat@height>\textheight\textheight\else\Gin@nat@height\fi}
\makeatother
\setkeys{Gin}{width=\maxwidth,height=\maxheight,keepaspectratio}
\usepackage{caption}
\usepackage{float}
\usepackage{subfig}
\ifxetex
  \usepackage[setpagesize=false, 
              unicode=false, 
              xetex]{hyperref}
\else
  \usepackage[unicode=true]{hyperref}
\fi
\hypersetup{breaklinks=true,
            bookmarks=true,
            pdfauthor={},
            pdftitle={WISE: full-Waveform variational Inference via Subsurface Extensions},
            colorlinks=true,
            citecolor=black,
            urlcolor=blue,
            linkcolor=black,
            pdfborder={0 0 0}}
\usepackage[preprint]{neurips_2019}
\usepackage{url}
\usepackage{booktabs}
\usepackage{amsfonts}
\usepackage{nicefrac}
\usepackage{microtype}

\makeatletter
\let\@oldfnsymbol\@fnsymbol
\renewcommand{\@fnsymbol}[1]{\@oldfnsymbol{0}}
\makeatother

\title{WISE: full-Waveform variational Inference via Subsurface Extensions}
\author{
    Ziyi Yin\textsuperscript{1,*}\thanks{* First two authors contributed equally. Correspondence: \href{mailto:ziyi.yin@gatech.edu}{ziyi.yin@gatech.edu}},
    Rafael Orozco\textsuperscript{1,*},
    Mathias Louboutin\textsuperscript{2},
    Felix J. Herrmann\textsuperscript{1}\\
    \textsuperscript{1}Georgia Institute of Technology,
    \textsuperscript{2}Devito Codes Ltd
}
\date{}

\begin{document}
\maketitle
\begin{abstract}
We introduce a probabilistic technique for full-waveform inversion,
employing variational inference and conditional normalizing flows to
quantify uncertainty in migration-velocity models and its impact on
imaging. Our approach integrates generative artificial intelligence with
physics-informed common-image gathers, reducing reliance on accurate
initial velocity models. Considered case studies demonstrate its
efficacy producing realizations of migration-velocity models conditioned
by the data. These models are used to quantify amplitude and positioning
effects during subsequent imaging.
\end{abstract}

\section{Introduction}\label{introduction}

Full-waveform inversion (FWI) plays a pivotal role in exploration,
primarily focusing on estimating Earth's subsurface properties from
observed seismic data \citep{virieux2009overview}. The inherent
complexity of FWI stems from its nonlinearity, further complicated by
ill-posedness and computational intensiveness of the wave modeling. To
address these challenges,~we introduce a computationally cost-effective
probabilistic framework that generates multiple migration-velocity
models conditioned on observed seismic data. By combining deep learning
\citep{yu2021deep} with physics, our approach harnesses advancements in
variational inference \citep[VI,][]{jordan1999} and generative
artificial intelligence
\citep[AI,][]{ramesh2022hierarchical, herrmann2023president}. We achieve
this by forming common-image gathers
\citep[CIGs,][]{symes2008migration}, followed by training conditional
normalizing flows \citep[CNFs,][]{winkler2019learning} that quantify
uncertainties in migration-velocity models.

Outline: First, we delineate the FWI problem and its inherent
challenges. Subsequently, we explore VI to quantify FWI's uncertainty.
To reduce VI's computational costs, we introduce \emph{physics-informed
summary statistics} and justify the use of CIGs as these statistics. Our
framework's capabilities are validated through two case studies, which
include studying the effects of uncertainty in the generated
migration-velocity models on migration.

\section{Methodology}\label{methodology}

We present a Bayesian inference approach to FWI by briefly introducing
FWI and VI used as a framework for uncertainty quantification (UQ).

\subsection{Full-waveform inversion}\label{full-waveform-inversion}

Estimation of unknown migration-velocity models, $\mathbf{x}$, from
noisy seismic data, $\mathbf{y}$ involves inverting nonlinear forward
modeling, $\mathcal{F}$, which links $\mathbf{x}$ to $\mathbf{y}$ via
$\mathbf{y} = \mathcal{F}(\mathbf{x}) + \boldsymbol{\epsilon}$ with
$\boldsymbol{\epsilon}$ measurement noise. Source/receiver signatures
are assumed known and absorbed in $\mathcal{F}$. Solving this nonlinear
inverse problem is challenging because of the noise and the non-trivial
null-space of the modeling \citep{tarantola1984}. As a result, multiple
migration-velocity models fit the data, necessitating a Bayesian
framework for UQ.

\subsection{Full-waveform inference}\label{full-waveform-inference}

Rather than seeking a single migration-velocity model, our goal is to
invert for a range of models compatible with the data, termed
``full-waveform inference''. From a Bayesian perspective, this involves
determining the posterior distribution of migration-velocity models
given the data, $p(\mathbf{x}|\mathbf{y})$. Algorithms to calculate
statistics from the posterior can be divided into: (i) sampling based
algorithms, such as MCMC
\citep{welling2011bayesian, ely2018assessing, kotsi2020uncertainty, siahkoohi2022};
or (ii) optimization based algorithms, such as VI \citep{jordan1999}.

Due to the high-dimensionality and expensive wave-based modeling, MCMC
becomes impractical for FWI \citep{gelman2003}. For this reason, we
focus on low-cost VI, which exchanges the computational cost of
posterior sampling for neural network training
\citep{rizzuti2020, ren2021seismic, kumar2021, siahkoohi2021, siahkoohi2021preconditioned, zhang2021introduction, siahkoohi2022wave, orozco2023adjoint, siahkoohi2023, wen2023conditional, orozco2023amortized, zhang20233, orozco2023refining, zhang2023bayesian, strutz2023variational, gahlot2023inference}.
Specifically, we employ amortized VI, which incurs offline computational
training cost but enables cheap online posterior inference on many
datasets $\mathbf{y}$ \citep{kruse2021hint}. Next, we discuss how to use
CNFs for amortized VI.

\subsection{Amortized variational inference with conditional normalizing
flows}\label{amortized-variational-inference-with-conditional-normalizing-flows}

During VI, the posterior distribution $p(\mathbf{x}|\mathbf{y})$ is
approximated by the surrogate,
$p_{\boldsymbol{\theta}}(\mathbf{x}|\mathbf{y})$, with learnable
parameters, $\boldsymbol{\theta}$. Because of their low-cost training
and rapid sampling \citep{rezende2015variational, louboutin2023learned},
CNFs are suitable to act as surrogates for the posterior with training
that involves minimization of the Kullback-Leibler divergence between
the true and surrogate posterior distribution. In practice, this
requires access to $N$ training pairs of migration-velocity model and
observed data to minimize the following objective:
\begin{equation}
\underset{\boldsymbol{\theta}}{\operatorname{minimize}} \quad \frac{1}{N}\sum_{i=1}^{N} \left(\frac{1}{2}\|f_{\boldsymbol{\theta}}\left(\mathbf{x}^{(i)};\mathbf{y}^{(i)}\right)\|_2^2-\log\left|\det\mathbf{J}_{f_{\boldsymbol{\theta}}}\right|\right).
\label{eq-cnf}
\end{equation}
 Here, $f_{\boldsymbol{\theta}}$ is the CNF with network parameters,
$\boldsymbol{\theta}$, and Jacobian,
$\mathbf{J}_{f_{\boldsymbol{\theta}}}$. It transforms each velocity
model, $\mathbf{x}^{(i)}$, into white noise (as indicated by the
$\ell_2$-norm), conditioned on the observation, $\mathbf{y}^{(i)}$.
After training, the inverse of CNF turn random realizations of the
standard Gaussian distribution into posterior samples
(migration-velocity models) conditioned on any seismic observation that
is in the same statistical distribution as the training data.

\subsection{Physics-informed summary
statistics}\label{physics-informed-summary-statistics}

While CNFs are capable of approximating the posterior distribution,
training the CNFs on pairs $\left(\mathbf{x},\,\mathbf{y}\right)$
presents challenges when changes in the acquisition occur or when
physical principles simplifying the mapping between model and data are
lacking, both of which lead to increasing training costs. To tackle
these challenges, \citet{radev2022} introduced fixed reduced-size
\emph{summary statistics} that encapsulate observed data and inform the
posterior distribution. Building on this concept,
\citet{orozco2023adjoint} uses the gradient as the
\emph{physics-informed summary statistics}, partially reversing the
forward map and therefore accelerating CNF training. For linear inverse
problems with Gaussian noise, these statistics are unbiased ---
maintaining the same posterior distribution, whether conditioned on
original data or on the gradient. Based on this principle,
\citet{siahkoohi2022wave} and \citet{siahkoohi2023} used reverse-time
migration \citep[RTM,][]{baysal1983reverse}, given by the action of the
adjoint of the linearized Born modeling
\citep{schuster1993, nemeth1999, alsing2018generalized}, to summarize
data and quantify imaging uncertainties for a fixed accurate
migration-velocity model.

We aim to extend this approach to the nonlinear FWI problems. While RTM
transfers information from the data to the image domain, its performance
diminishes for incorrect migration velocities. \citet{hou2016} showed
that least-squares migration \citep{zeng2014} can perfectly fit the data
for correct migration-velocity models, but this fit fails for wrong
velocity models. This highlights a fundamental limitation in cases where
the velocity model is wrong and RTM does not capture the information,
which leads to a biased posterior. For a wrong initial FWI-velocity
model $\mathbf{x}_0$,
$p\left(\mathbf{x}\middle|\mathbf{y}\right) \neq p\left(\mathbf{x}\middle|\nabla\mathcal{F}\left(\mathbf{x}_0\right)^\top \mathbf{y}\right)$
with $\nabla\mathcal{F}$ Born modeling and $^\top$ the adjoint. To avoid
this problem, more robust \emph{physics-informed summary statistics} are
needed to preserve information.

\subsection{Common-image gathers as summary
statistics}\label{common-image-gathers-as-summary-statistics}

Migration-velocity analysis has a rich history in the literature
\citep{sava2004wave, symes2008migration}. Following
\citet{liu2013multisource};\citet{hou2015approximate};\citet{hou2018inversion},
we employ relatively artifact-free subsurface-offset extended Born
modeling to calculate summary statistics. Thanks to being closer to an
isometry---i.e, the adjoint of extended Born modeling is closer to its
inverse \citep{yang2021, kroode2023} and therefore preserves
information, its adjoint can nullify residuals even when the
FWI-velocity model is incorrect as shown by \citet{hou2016}.
\citet{geng2022deep} further demonstrated that neural networks can be
used to map CIGs to velocity models. Both these findings shed important
light on the role of CIGs during VI because CIGs preserve more
information, which leads to less biased physics-informed summary
statistics for a wrong initial FWI-velocity model. Formally, this means
$p\left(\mathbf{x}\middle|\mathbf{y}\right) \approx p\left(\mathbf{x}\middle|\overline{\nabla\mathcal{F}}\left(\mathbf{x}_0\right)^\top \mathbf{y}\right)$,
where $\overline{\nabla\mathcal{F}}$ is extended Born modeling.
Leveraging this mathematical observation, we propose WISE, short for
full-\textbf{W}aveform variational \textbf{I}nference via
\textbf{S}ubsurface \textbf{E}xtensions. The core of this technique is
to train CNFs with pairs of velocity models, $\mathbf{x}$, and CIGs,
$\overline{\nabla\mathcal{F}}\left(\mathbf{x}_0\right)^\top \mathbf{y}$,
guided by the objective of Equation~\ref{eq-cnf}. Our case studies will
demonstrate that even with wrong initial FWI-velocity models, CIGs
encapsulate more information, enabling the trained CNFs to generate
accurate migration-velocity models conditioned by the data.

\section{Synthetic case studies}\label{synthetic-case-studies}

Our study evaluates the performance of WISE through synthetic case
studies on two datasets: the CurveFault-A dataset of Open FWI
\citep{deng2022openfwi} and 2D slices of the Compass dataset
\citep{e.jones2012}. We aim to compare the quality of posterior samples
informed by RTM alone versus those informed by CIGs.

\subsection{Open FWI}\label{open-fwi}

The CurveFault-A dataset comprises velocity models with significant
variability across samples, which poses challenges for deep learning
methods \citep{jin2021unsupervised, jin2023does}. This is further
compounded by faults and dipping events while observations contain only
reflected energy. Testing on this dataset allows us to test WISE's
velocity-model generation capabilities.

\textbf{Dataset generation and network training.} We select $3000$
velocity models of $640$ m by $640$ m, each with $64$ equally spaced
receivers at $10\mathrm{m}$ tow depth and $16$ randomly placed sources
\citep{herrmann2010randomized}. The surface is assumed absorbing. Using
a $15\mathrm{Hz}$ central frequency Ricker wavelet with energy below
$3\mathrm{Hz}$ removed for realism, acoustic data is simulated with
Devito \citep{devito-api, devito-compiler} and JUDI.jl
\citep{witteJUDI2019, judi}. Uncorrelated band-limited Gaussian noise is
added (S/N $12\mathrm{dB}$) before migrating each dataset with a 1D
initial FWI-velocity model calculated by averaging the corresponding
true model horizontally. CIGs are computed for $101$ subsurface offsets
ranging from $-250\mathrm{m}$ to $+250\mathrm{m}$
\citep{louboutin2023, louboutin2023a}. The dataset is split into $2800$
training, $150$ validation, and $50$ test samples. Two CNFs are trained:
one with velocity-RTM pairs and another with velocity-CIGs pairs.

\begin{figure}
\centering
\includegraphics[width=0.990\hsize]{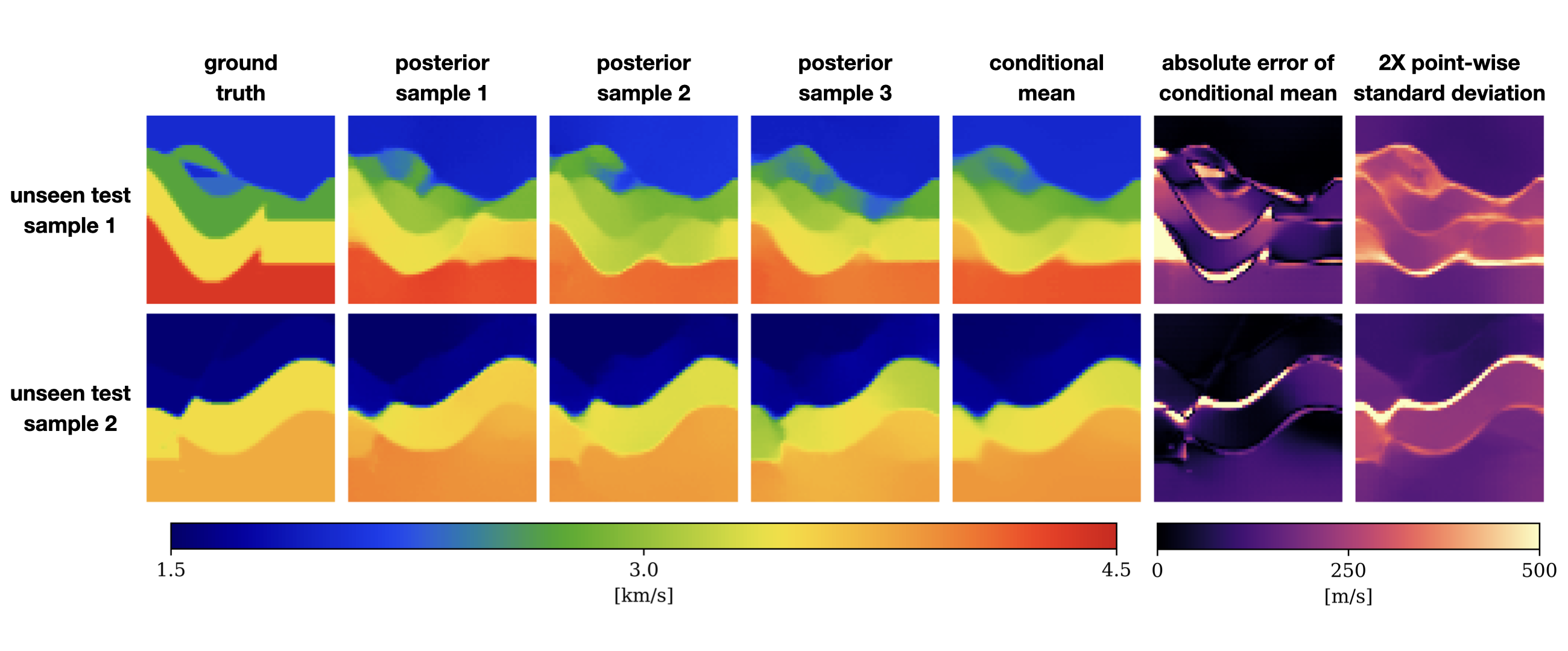}
\caption{Applying WISE for two unseen test samples in Open FWI
CurveFault-A dataset.}\label{fig-open-fwi}
\end{figure}

\textbf{Results.} Results on two untested samples by our CNFs are
included in Figure~\ref{fig-open-fwi} and reveal notable variation in
the posterior samples for sharp boundaries and smooth transitions in the
velocity. While the conditional mean estimate does not fully replicate
the true velocity, the standard deviations meaningfully correlate with
the errors, indicating that the uncertainty represented by the standard
deviation is informative. Across $50$ test samples, the mean SSIM score
for CIGs-based statistics is $0.87$, surpassing the $0.85$ mean for
RTM-based statistics. Motivated by these results, we will examine a more
realistic example with intricate geological structures next.

\subsection{Compass model}\label{compass-model}

To validate WISE in a more realistic setting and examine uncertainty in
imaging, we consider the Compass dataset, known for its ``velocity
kickback'' challenge for FWI algorithms. By comparing, for a poor
initial FWI-velocity model, the conditional mean of samples of the
migration-velocity model informed by CIGs and RTM, we verify the
superior information content of CIGs. We also illustrate how uncertainty
in migration-velocity models can be converted into uncertainties in
amplitude and positioning of imaged reflectors
\citep{poliannikov2016effect}.

\begin{figure}
\centering
\captionsetup[subfigure]{labelformat=empty}
\subfloat[(a)]{\includegraphics[width=0.490\hsize]{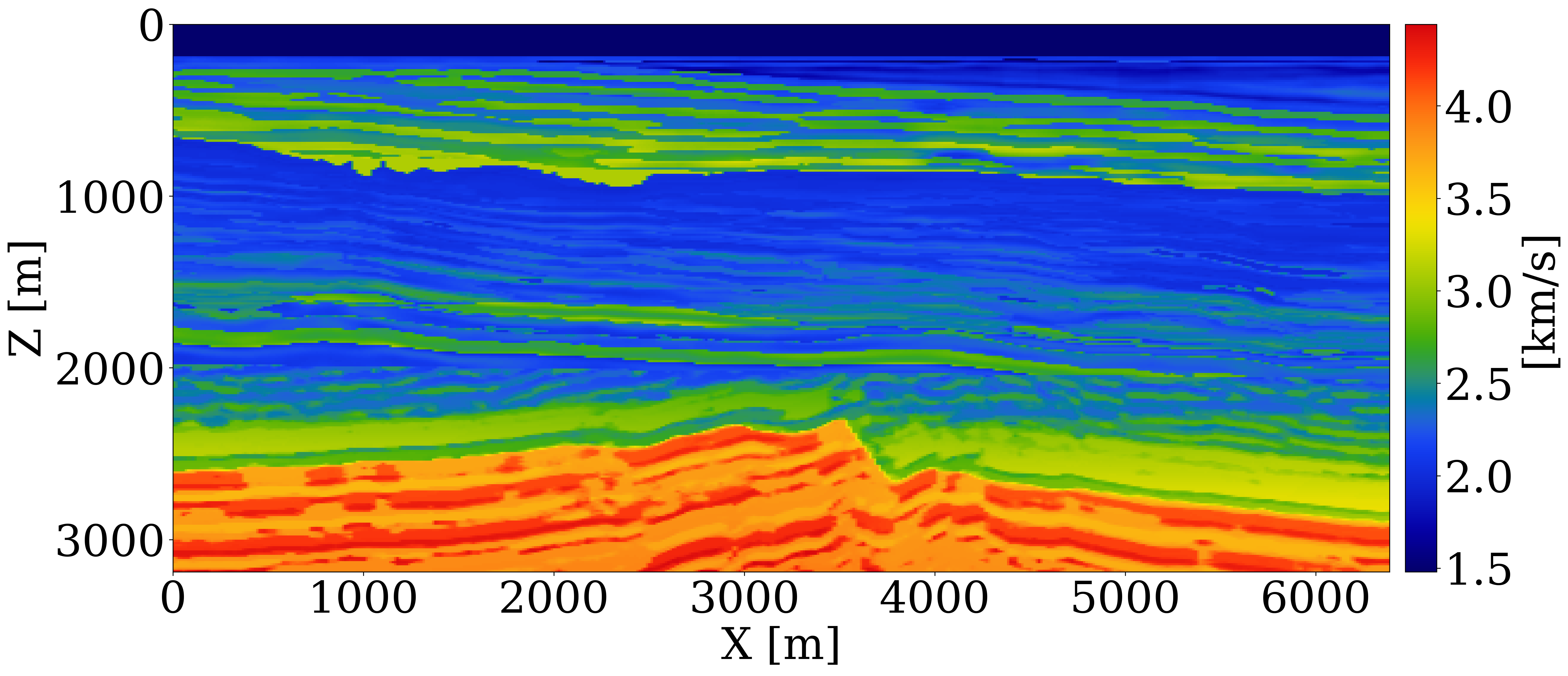}}
\subfloat[(b)]{\includegraphics[width=0.490\hsize]{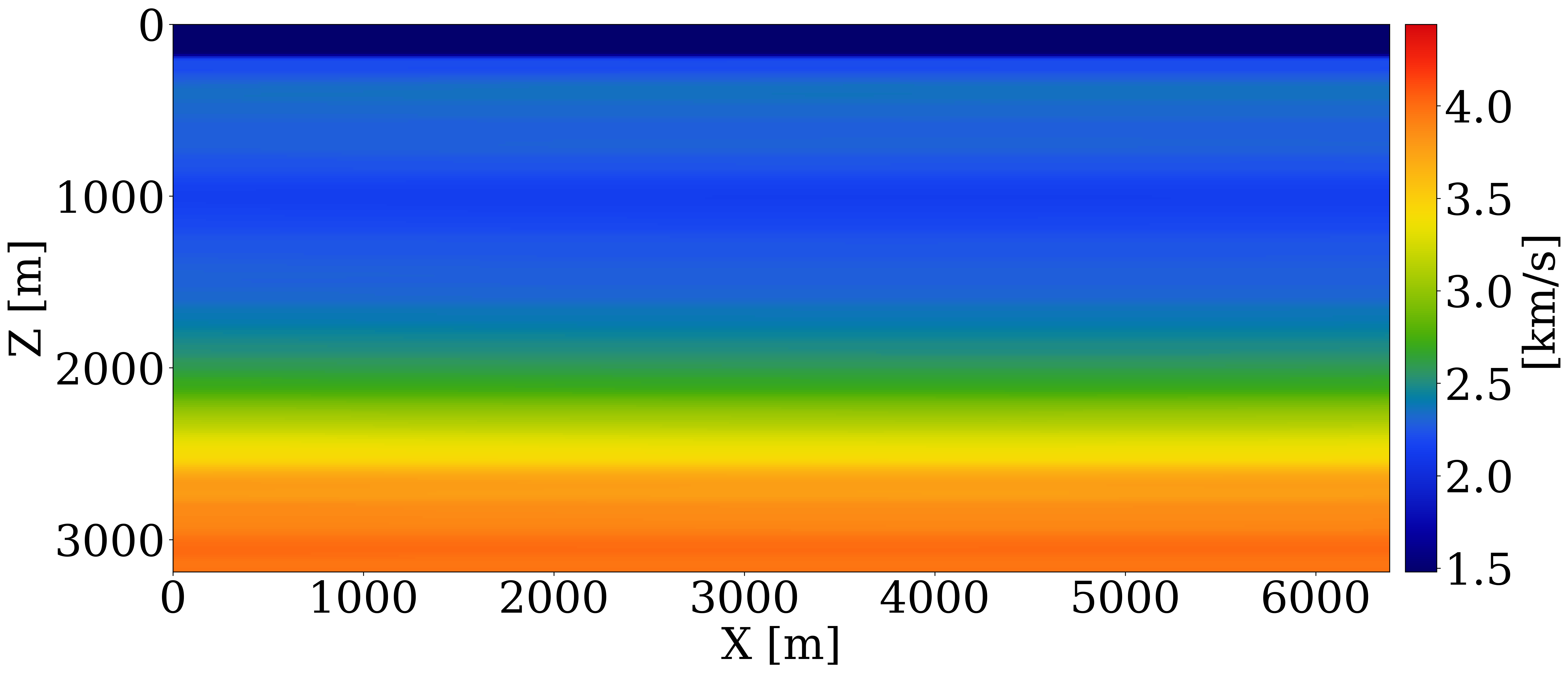}}
\\
\subfloat[(c)]{\includegraphics[width=0.490\hsize]{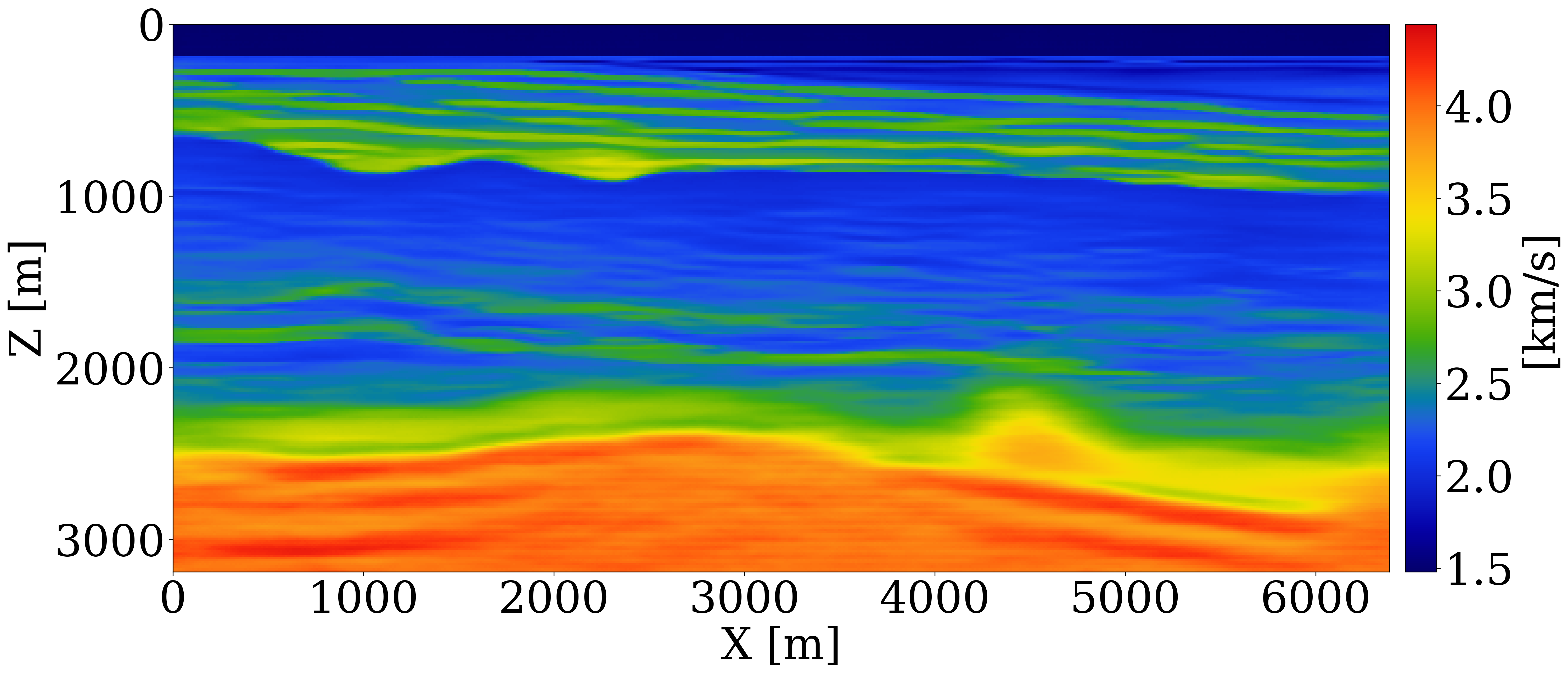}}
\subfloat[(d)]{\includegraphics[width=0.490\hsize]{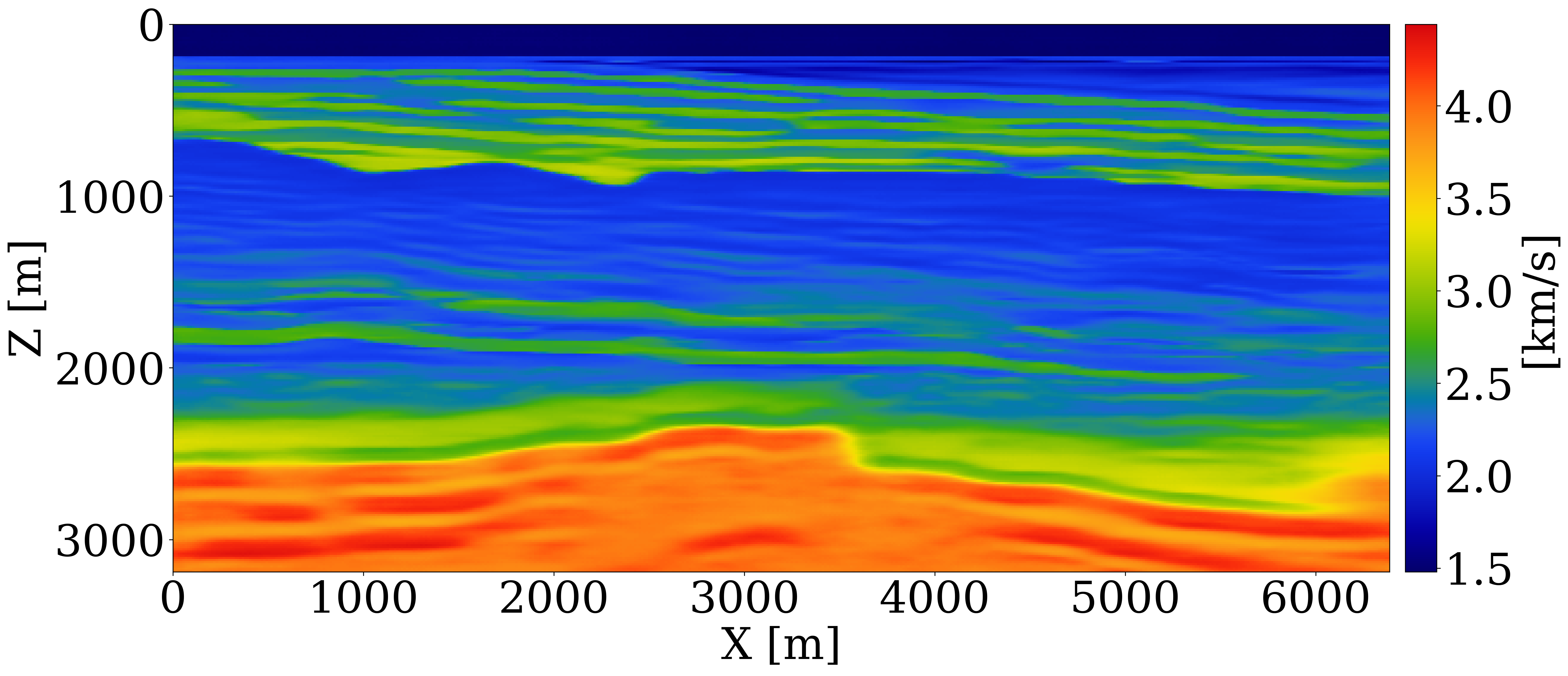}}
\caption{(a) an unseen ground-truth velocity model; (b) 1D initial
FWI-velocity model; (c) conditional mean estimate for RTM as summary
statistics ($\mathrm{SSIM}=0.48$); (d) conditional mean estimate from
WISE ($\mathrm{SSIM}=0.56$).}\label{fig-true-migration}
\end{figure}

\begin{figure}
\centering
\captionsetup[subfigure]{labelformat=empty}
\subfloat[(a)]{\includegraphics[width=0.490\hsize]{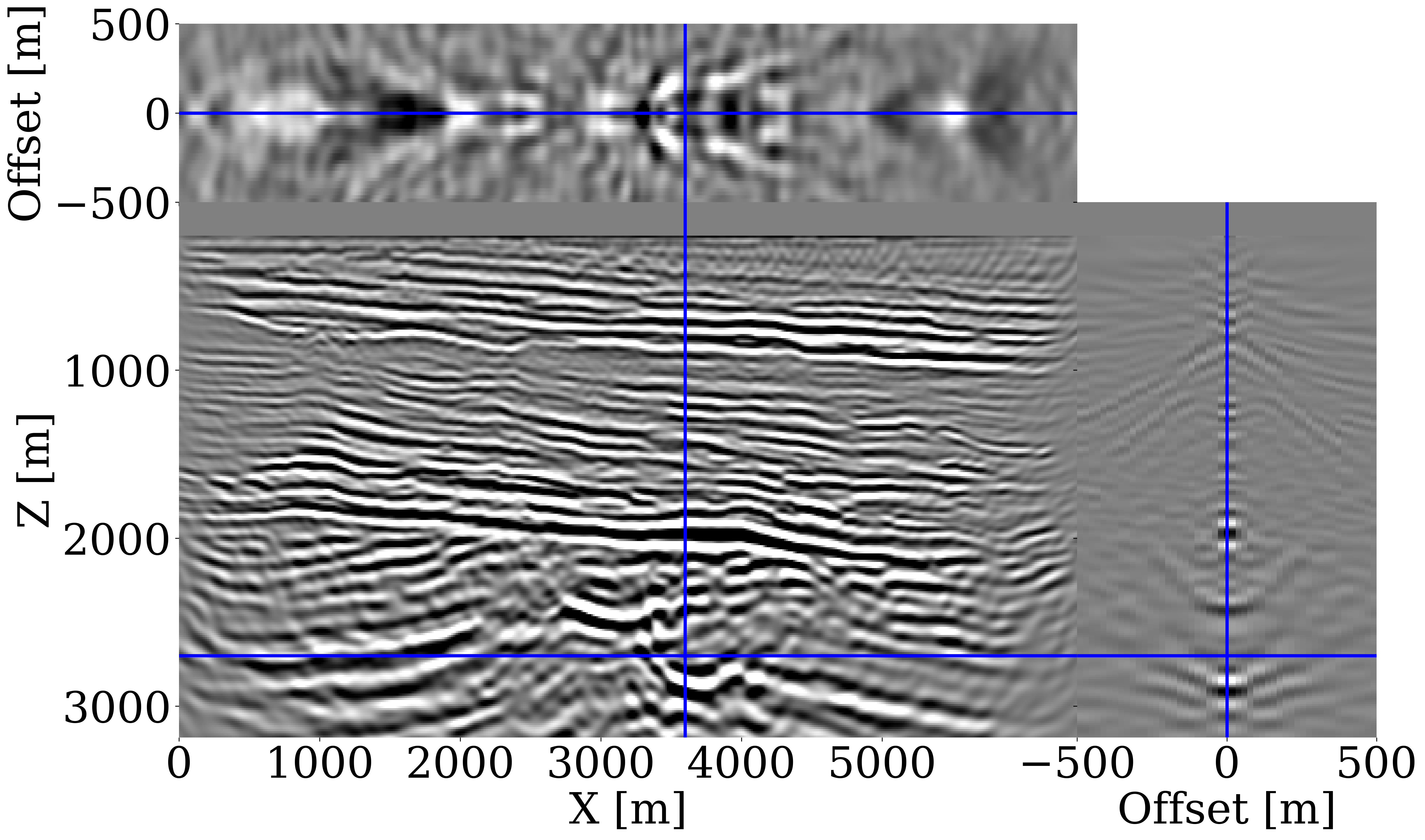}}
\subfloat[(b)]{\includegraphics[width=0.490\hsize]{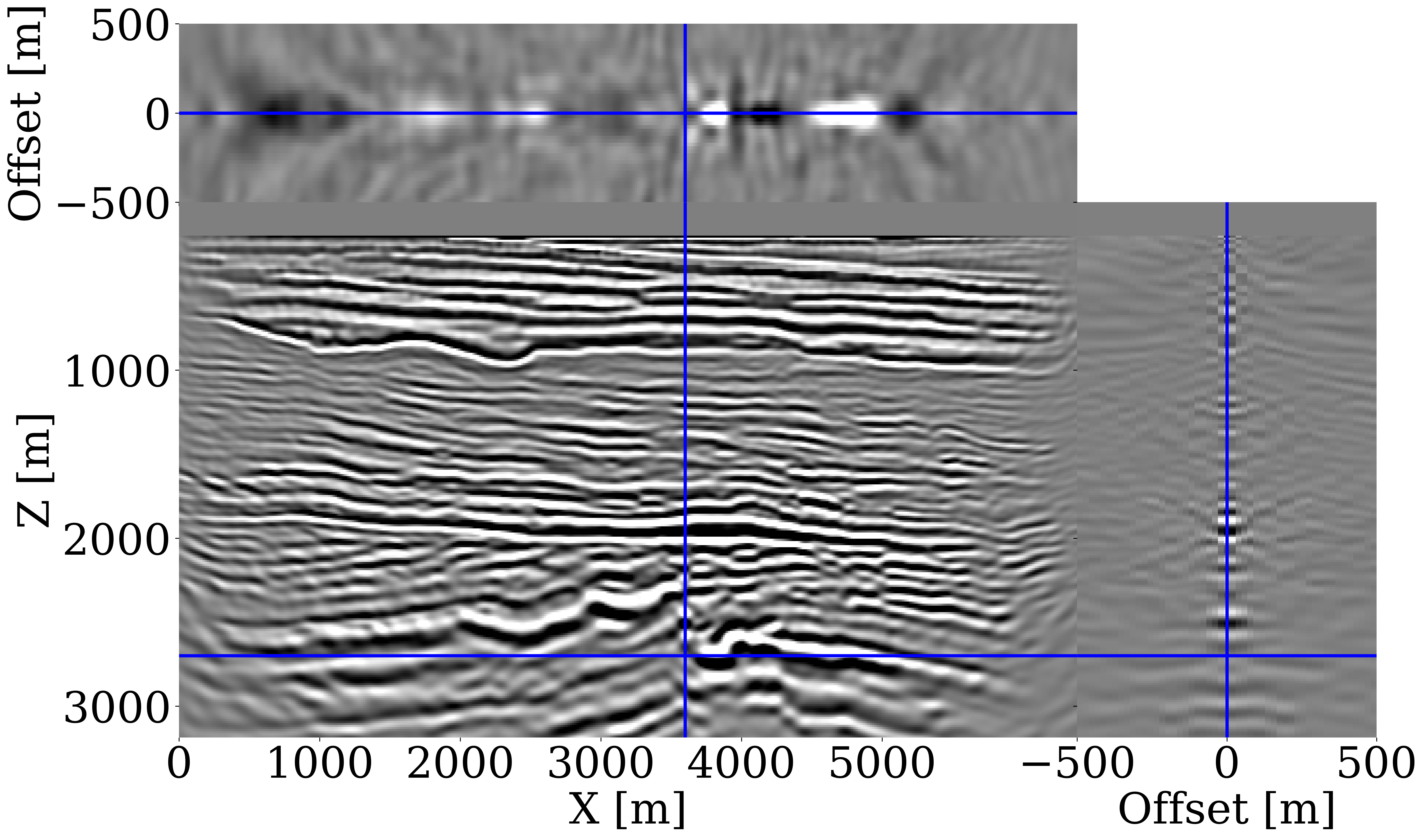}}
\caption{CIGs calculated by the initial FWI-velocity model given by (a)
the 1D background model (shown in Figure~\ref{fig-true-migration}(b)) or
by (b) the conditional mean estimate (shown in
Figure~\ref{fig-true-migration}(d)).}\label{fig-qc}
\end{figure}

\textbf{Dataset generation and network training.} We take $1040$ 2D
slices of the Compass model of $6.4$ km by $3.2$ km, with $512$ equally
spaced sources towed at $12.5\mathrm{m}$ depth and $64$ ocean-bottom
nodes at random locations. Source and noise setup remains the same. The
arithmetic mean over all velocity models is used as the 1D initial
FWI-velocity model (shown in Figure~\ref{fig-true-migration}(b)). $51$
subsurface offsets ranging from $-500\mathrm{m}$ to $+500\mathrm{m}$ are
used to compute CIGs (shown in Figure~\ref{fig-qc}(a)). The dataset is
divided into $800$ training, $190$ validation, and $50$ test samples,
with CNFs trained over $200$ epochs.

\textbf{Results.} Our method's performance is evaluated on an unseen 2D
Compass slice shown in Figure~\ref{fig-true-migration}(a). When RTM is
used to summarize the data, the conditional mean estimate
(Figure~\ref{fig-true-migration}(c)) does not capture the shape of the
unconformity. Thanks to the CIGs, WISE captures more information and as
a result produces a more accurate conditional mean
(Figure~\ref{fig-true-migration}(d)). For the $50$ test samples, the
SSIM scores with CIGs yield a mean of $0.63$, outperforming RTM-based
statistics with a mean SSIM of $0.52$.

\textbf{Quality control.} To verify the inferred migration-velocity
model, CIGs calculated for the initial FWI-velocity model
(Figure~\ref{fig-true-migration}(b)), plotted in Figure~\ref{fig-qc}(a),
are juxtaposed against CIGs calculated for the inferred
migration-velocity model (Figure~\ref{fig-true-migration}(d)), plotted
in Figure~\ref{fig-qc}(b). Significant improvement in near-offset
focused energy is observed in the CIGs for the inferred
migration-velocity model. A similar focusing behavior is noted for the
posterior samples themselves.

\begin{figure}
\centering
\captionsetup[subfigure]{labelformat=empty}
\subfloat[(a)]{\includegraphics[width=0.490\hsize]{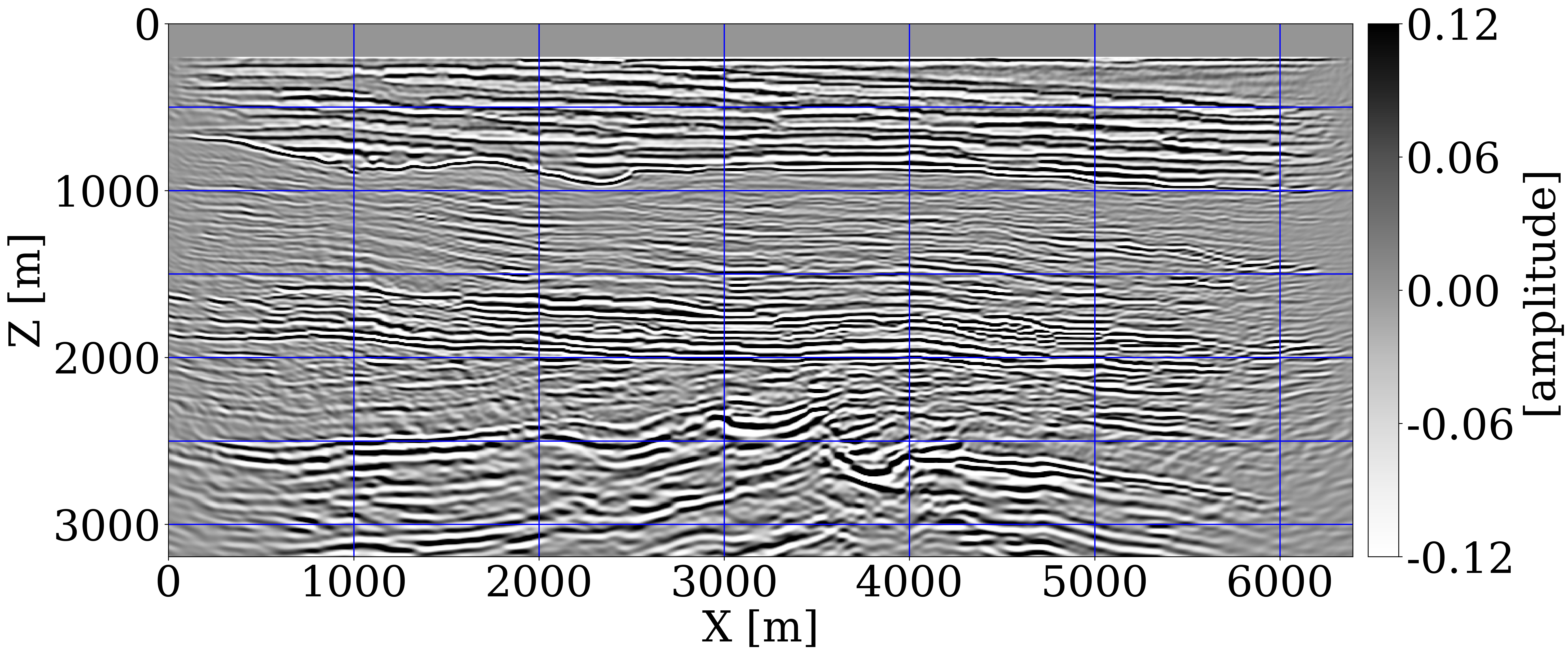}}
\subfloat[(b)]{\includegraphics[width=0.490\hsize]{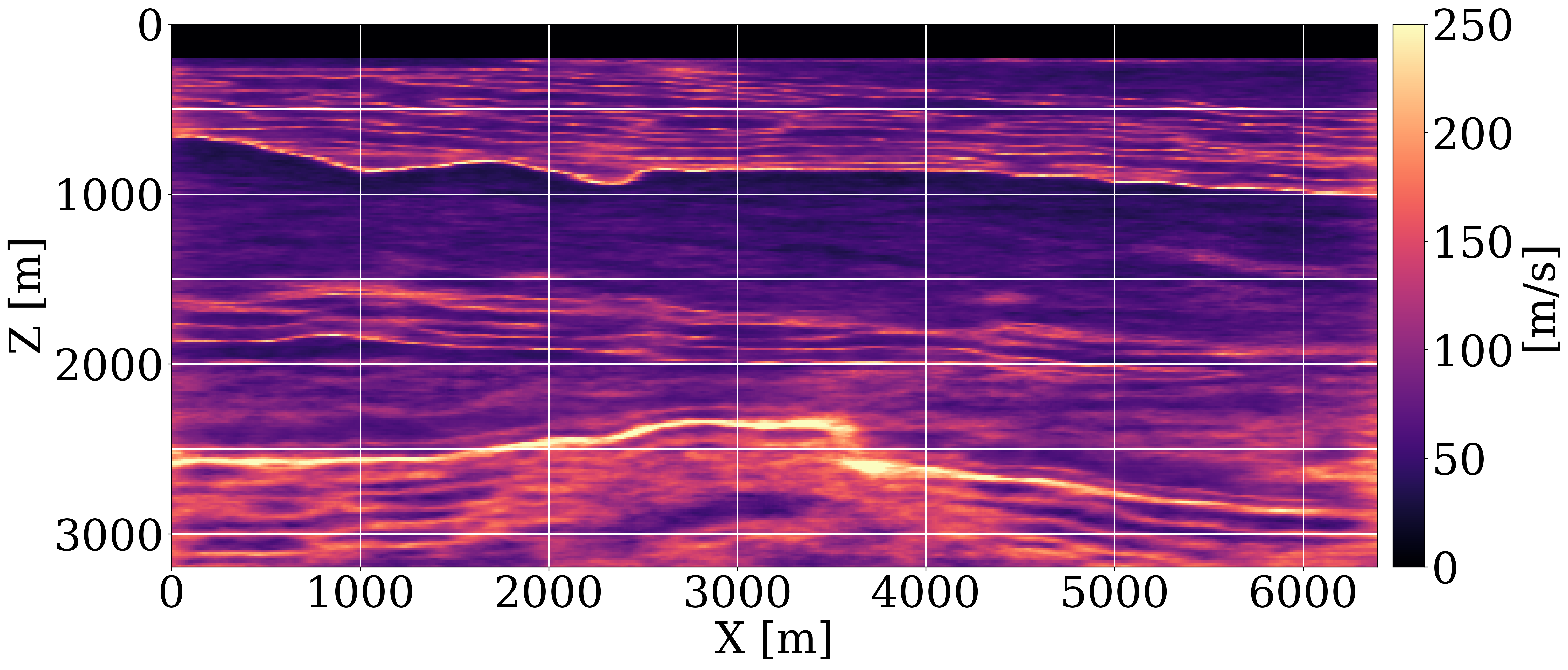}}
\\
\subfloat[(c)]{\includegraphics[width=0.490\hsize]{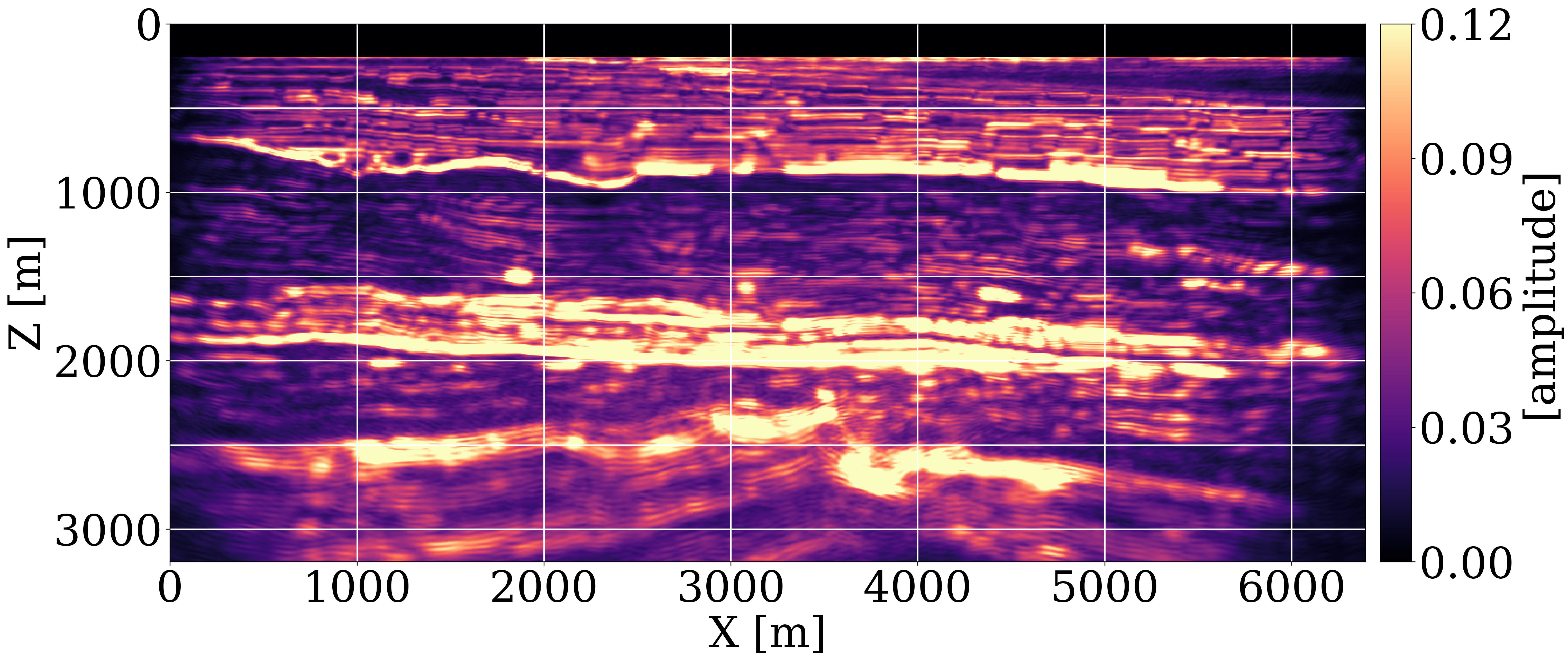}}
\subfloat[(d)]{\includegraphics[width=0.490\hsize]{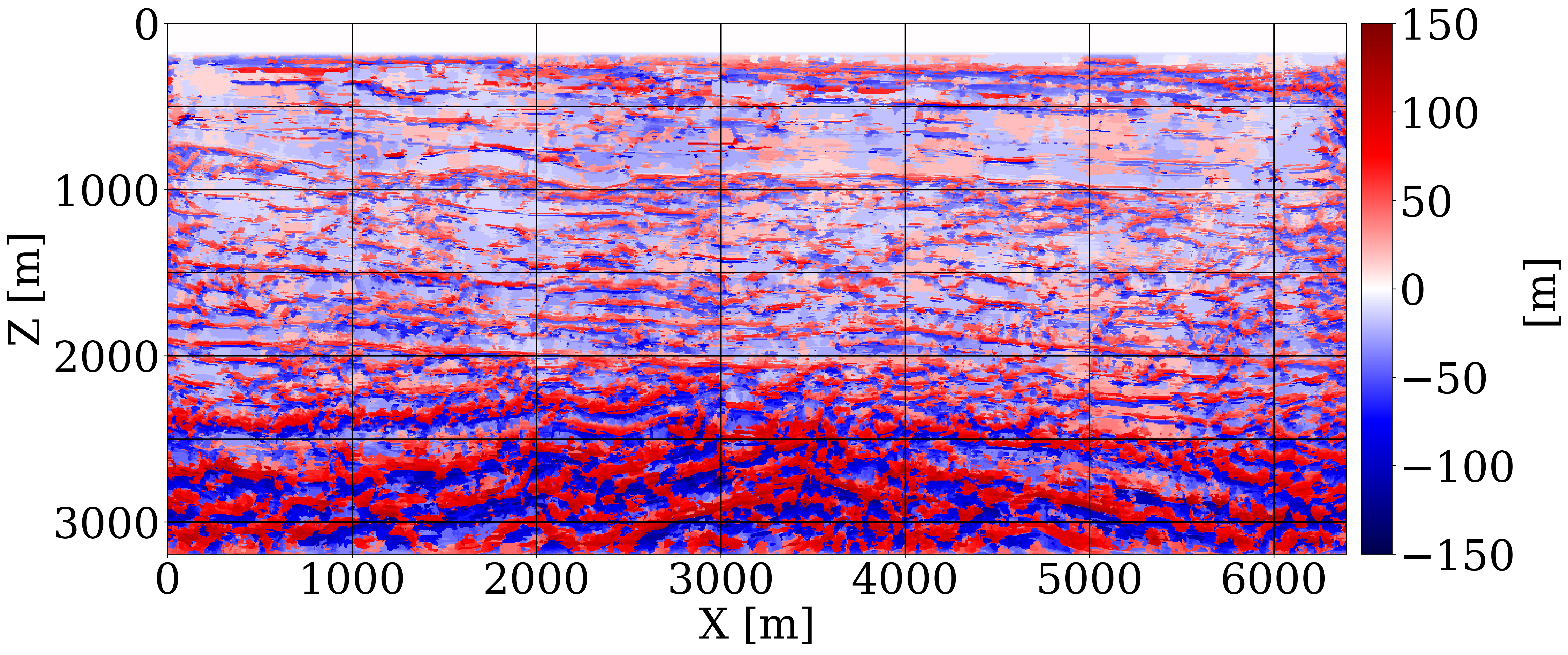}}
\caption{Using WISE in the downstream imaging task: (a) imaged
reflectors using the conditional mean estimate (shown in
Figure~\ref{fig-true-migration}(d)) as the migration velocity; (b)
point-wise standard deviation of the posterior velocity samples; (c)
point-wise deviation of the imaged reflectivities; (d) point-wise
maximum depth shift. Note: (a) and (c) are normalized by the same
constant. The images are all grided for visualization
purpose.}\label{fig-post}
\end{figure}

\textbf{Downstream imaging.} While access to the posterior represents an
important step towards grasping uncertainty, understanding its impact on
imaging with ($30\mathrm{Hz}$) RTMs is more relevant because it concerns
uncertainty in the final product. For this purpose, comparisons are made
between the RTM computed for the conditional mean
(Figure~\ref{fig-post}(a)) and RTMs computed for individual posterior
samples that vary significantly judged by the point-wise standard
deviations in Figure~\ref{fig-post}(b). These deviations increase with
depth and correlate with complex geology where the RTM-based inference
struggled. To understand how this uncertainty propagates to RTM images,
forward uncertainty is assessed by carrying out RTMs for different
posterior samples with results for the standard deviations plotted in
Figure~\ref{fig-post}(c). These amplitude deviations are different
because mapping migration-velocities to RTMs is highly nonlinear,
leading to large areas of intense amplitude variation and dimming at the
edges caused by the Born modeling's null-space. While these amplitude
sensitivities are useful, deviations in the migration velocities also
leads to differences in reflector positioning. Vertical shifts between
the reference image (Figure~\ref{fig-post}(a)) and RTMs for different
posterior samples are calculated with a local cross-correlation
technique \citep{hale2006efficient} and included in
Figure~\ref{fig-post}(d) where blue/red areas correspond to up/down
shifts. As expected, these shifts are most notable in the deeper regions
and at the edges where velocity variations are the largest.

\section{Discussion}\label{discussion}

Once the offline costs of computing 800 CIGs and network training are
covered, WISE enables generation of velocity models for unseen seismic
data at the low computational cost of a single set of CIGs for a poor
initial FWI-velocity model. The Open FWI case study demonstrates WISE's
capability of producing realistic posterior samples and conditional
means for a broad range of unseen velocity models. In the case of the
Compass model, the initial FWI-velocity model was poor. Still, CIGs
obtained from a single 1D initial model capture relevant information
from the non-zero offsets. From this information, the network learns to
produce migration-velocity models that focus at inference. WISE also
produced two types of uncertainty, namely (i) inverse uncertainty in
migration-velocity model estimation from noisy data, and (ii) forward
uncertainty where uncertainty in migration-velocity models is propagated
to uncertainty in amplitude and positioning of imaged reflectors. The
latter can also be assessed for tasks like horizon tracking
\citep{wu2018least, siahkoohi2022} and seismic interpretation
\citep{kaur2023deep}.

Opportunities for future research remain. One area concerns dealing with
the ``amortization gap'' where CNFs tend to maximize performance across
multiple datasets rather than excelling at a single observation
\citep{marino2018iterative}. To improve single-observation performance,
particularly for out-of-distribution samples, computationally more
expensive latent space corrections \citep{siahkoohi2023} can be employed
that incorporate the physics. Moreover, velocity continuation methods
\citep{fomel2003time, yang2021} could be used including recent advances
in neural operators \citep{siahkoohi2022velocity}. These could offset
the cost of running RTMs for each posterior sample.

\section{Conclusions}\label{conclusions}

We present WISE, full-\textbf{W}aveform variational \textbf{I}nference
via \textbf{S}ubsurface \textbf{E}xtensions, for computationally
efficient uncertainty quantification of FWI. This framework underscores
the potential of generative AI in addressing FWI challenges, paving the
way for a new seismic inversion and imaging paradigm that is
uncertainty-aware. By having common-image gathers act as
information-preserving summary statistics, a principled approach to UQ
is achieved where generative AI is successfully combined with wave
physics. Because WISE automatically produces distributions for
migration-velocity models conditioned by the data, it moves well beyond
traditional velocity model building. It was shown that this
distributional information can be employed to quantify uncertainties in
the migration-velocity models that can be used to better understand
amplitude and positioning uncertainty in migration.

\section{Acknowledgement}\label{acknowledgement}

This research was carried out with the support of Georgia Research Alliance and partners of the ML4Seismic Center. The authors would like to thank Charles Jones (Osokey) for the constructive discussion. The authors gratefully acknowledge the contribution of OpenAI's ChatGPT for refining sentence structure and enhancing the overall readability of this manuscript.

\bibliography{paper}

\end{document}